\documentstyle[aps,prl,subfigure,epsfig,amsmath,amssymb,twocolumn]{revtex}

\newcommand{\boldvec}[1]{\mbox{\boldmath$#1$}}
\newcommand{\smallvec}[1]{\mbox{\boldmath$\scriptstyle#1$}}
\newcommand{\annhilate}[2]{{#1}^{\phantom\dag}_{{#2}}}
\newcommand{\create}[2]{{#1}^{\dag}_{{#2}}}

\begin{document}

\draft

\title{Signatures of resonance superfluidity in a quantum Fermi gas }

\author{M. L. Chiofalo$^*$, S. J. J. M. F.  Kokkelmans, J. N.
  Milstein, and M. J. Holland}

\address{JILA, University of Colorado and National Institute of
  Standards and Technology, Boulder, Colorado 80309-0440}

\wideabs{

\maketitle                

\begin{abstract}
  In this letter, we predict a direct and observable signature of the
  superfluid phase in a quantum Fermi gas, in a temperature regime
  already accessible in current experiments. We apply the theory of
  resonance superfluidity to a gas confined in a harmonic potential
  and demonstrate that a significant increase in density will be
  observed in the vicinity of the trap center.
\end{abstract}


}

Following the successful realization of Bose-Einstein condensation
(BEC) in confined vapors~\cite{bec}, it is natural to consider
possibilities for observing the analogous superfluid phase transition
in a dilute Fermi gas. Quantum degeneracy has already been
demonstrated in a two-component Fermi gas of $^{40}$K
atoms~\cite{dfg}, although the spin states utilized were not suitable
for exhibiting superfluidity.  The lowest temperatures achieved to
date in this system are around $0.2\; T_F$---limited by Pauli blocking
as well as a number of technical considerations~\cite{holland}. In
other experiments, the rethermalization of fermion atoms by elastic
collisions with a bath of ultracold bosons is exploited; realized by
mixtures of $^6$Li and $^7$Li at Rice~\cite{dfghul}, and at
ENS~\cite{dfgparis}, and more recently by mixtures of $^{40}$K atoms
and $^{87}$Rb atoms at JILA~\cite{dfgJILA}.

In order to observe a superfluid phase transition at critical
temperatures as high as $0.2\,T_F$, the existence of a strong coupling
mechanism which could lead to a significant amount of Cooper pairing
is necessary. Several theoretical papers have presented models to
investigate this regime, essentially based on application of the
Bardeen-Cooper-Schrieffer (BCS)~\cite{bcs} theory of
superconductivity. These approaches consider dilute Fermi vapors in
which the two body scattering processes are characterized by a large
negative scattering length $a$~\cite{stoof}.  Under such conditions
the relevant length scale---the spatial extent of the Cooper
pair---may become comparable to the average interparticle spacing.
This places the system in a crossover region from the BCS
superfluidity of momentum-correlated fermion pairs to the BEC of
tightly bound composite bosons. In this crossover regime, fluctuations
play a crucial role~\cite{crossover} and must be addressed.

Eventually, as the coupling is increased, it becomes necessary to
construct a theory in which explicit treatment of the composite
bosonic states is made. Such an approach was proposed in the context
of high-temperature superconductivity~\cite{lee} and is based on an
effective many-body Hamiltonian, in which quasibound pairs are
explicitly treated as resonance states embedded in the continuum of
the Fermi sea.  Such resonances are ubiquitous in atomic physics,
where, for example, a Feshbach resonance~\cite{feshbach} can be
utilized to tune a quasibound state through threshold, providing an
explicit microscopic basis for a theory of resonance
superfluidity~\cite{PRL,timmermans}.

A convincing method for detecting the superfluidity will be required.
Various approaches have been proposed; including measurements of the
pair distribution~\cite{static}, experiments involving the breakup of
the Cooper pairs~\cite{torma}, measurements of the moment of
inertia~\cite{vinas}, and probes of collective
excitations~\cite{minguzzi,baranov}. In this letter, we show that a
more straightforward and direct experimental signature of the
transition to the superfluid phase is provided by the density
characteristics in an inhomogeneous system. We demonstrate that in a
harmonic trap, the superfluid state is manifest as the appearance of a
bulge in the central atomic density. To this aim we derive a theory of
resonance superfluidity including the description of external
confinement.

\begin{figure}[h]
\begin{center}\
  \epsfysize=60mm \epsfbox{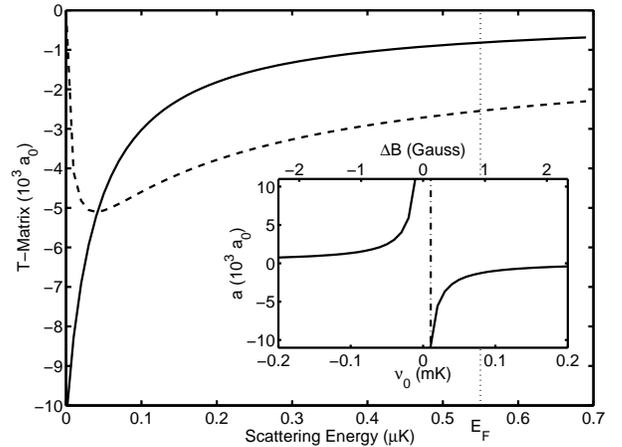}
\end{center}
\caption{
  Real (solid line) and imaginary (dashed line) components of the
  $T$-matrix for collisions of the lowest two spin states of $^{40}$K
  at a detuning of 20~$E_F$, shown in length dimensions, i.e.\ 
  $T_k/(4\pi\hbar^2/m)$. The scattering length is the intercept at
  zero scattering energy which for this case is approximately
  -10000~$a_0$, where $a_0$ is the Bohr radius. The large variation in
  the $T$-matrix over the relevant energy range indicates that a
  quantum field theory developed from this microscopic basis will in
  general need to account for physics beyond the scattering length
  approximation. The inset shows the scattering length as a function
  of detuning, with 20~$E_F$ detuning indicated by the dashed-dot
  line. This curve obeys the following form $a=a_{\rm
    bg}(1-\kappa/\nu_0)$ where $a_{\rm bg}=176\,a_0$ and
  $\kappa=0.657\,$mK~\protect\cite{bohn}. The quasipotentials to be
  renormalized are then $U_0=4\pi\hbar^2a_{\rm bg}/m$ and
  $g_0=\sqrt{\kappa U_0}$.}
\label{fig1}
\end{figure}

Although our results will have general applicability, for the purpose
of illustration, we consider a typical system of $N=5\times10^5$ atoms
in an isotropic harmonic trap with angular frequency
$\omega=2\pi\,100$~s$^{-1}$. This gives a Fermi energy of
$E_F=(3N)^{1/3}\hbar\omega$. We consider a $^{40}$K Feshbach
resonance, illustrated in Fig.~1, for $s$-wave scattering of atoms in
the lowest two hyperfine spin states which we denote by
$\sigma\in\{\uparrow,\downarrow\}$. We begin by considering the
general structure of the theory for the homogeneous system where the
fermions are represented by the wave-number-$\boldvec k$ dependent
annihilation operators $\annhilate{a}{\smallvec k\sigma}$ and the
composite boson field by $b_{\smallvec k}$. The Hamiltonian is
\begin{eqnarray} 
  H&=&\sum_{\smallvec k\sigma}\epsilon_k\, \create{a}{\smallvec
    k\sigma} \annhilate{a}{\smallvec k\sigma} +\nu\,\sum_{\smallvec
    k}\create{b}{\smallvec k} \annhilate{b}{\smallvec k} \nonumber\\
  &&{}+U\sum_{\smallvec q\smallvec k\smallvec k'} \create{a}{\smallvec
    q/2+\smallvec k\uparrow}\create{a}{\smallvec q/2-\smallvec k
    \downarrow} \annhilate{a}{\smallvec q/2-\smallvec k'
    \downarrow}\annhilate{a}{\smallvec q/2+\smallvec k'\uparrow}
  \nonumber\\
  &&{}+\Bigl(g\sum_{\smallvec k\smallvec q} \create{b}{\smallvec q}
  \annhilate{a}{\smallvec q/2-\smallvec k
    \downarrow}\annhilate{a}{\smallvec q/2+\smallvec k
    \uparrow}+\mbox{h.c.}\Bigr)\,,
\end{eqnarray}
where h.c.\ denotes the hermitian conjugate. The free dispersion
relation for the fermions is $\epsilon_k$, and $\nu$ denotes the
detuning of the boson resonance state from the zero edge of the
collision continuum. The collisional interactions include both
background fermion scattering and interconversion between composite
bosons and fermion pairs. It is implicit in treating $U$ and $g$ as
constants that the theory will be renormalized, and thereby contain no
ultraviolet divergences in the calculation of observable
quantities~\cite{followingpaper}. This procedure involves ascribing a
cutoff value $K$ as the upper limit of all momentum summations, and
renormalizing the Hamiltonian constants in terms of $K$ and the
parameters for the Feshbach resonance as given in Fig.~1. Defining
$\alpha=mK/(2\pi^2\hbar^2)$ and a dimensionless factor
$\Gamma=(1-\alpha U_0)^{-1}$, the renormalization is executed by the
following relations $U=\Gamma U_0$, $g=\Gamma g_0$, and $\nu=\nu_0+
\alpha gg_0$. All results presented here have been shown to be
independent of $K$.

From this Hamiltonian, we construct the dynamical
Hartree-Fock-Bogoliubov equations for both the bosonic and fermionic
mean-fields. These equations involve the mean fields corresponding to
the spin density $n=\sum_{\smallvec k}\langle\create{a}{\smallvec
  k\sigma}\annhilate{a}{\smallvec k\sigma}\rangle$ (taken to be
identical for both spins), the pairing field $p=\sum_{\smallvec
  k}\langle\annhilate{a}{-\smallvec
  k\downarrow}\annhilate{a}{\smallvec k\uparrow}\rangle$, and the
condensed boson field $\phi_m=\langle b_{\smallvec k=0}\rangle$. The
single particle density matrix
\begin{equation}
  {\cal G}_{i,j}=\langle \create{A}{j}\annhilate{A}{i}\rangle\,,\qquad
  A=\left(\begin{array}{c} \annhilate{a}{\smallvec k\uparrow}\\
      \annhilate{a}{\smallvec k\downarrow}\\ \create{a}{-\smallvec
        k\uparrow}\\ \create{a}{-\smallvec k\downarrow}
  \end{array} \right)
\end{equation}
evolves according to the Bogoliubov self-energy $\Sigma$
\begin{equation}
  i\hbar\frac{d{\cal G}}{dt} = [\Sigma,{\cal G}]\,.
\end{equation}
The self-energy has hermitian structure
\begin{equation}
  \Sigma=\left(\begin{array}{cccc}
      U_k & 0 & 0 & \Delta\\
      0 & U_k & -\Delta & 0\\
      0 & -\Delta^* & -U_k & 0\\
      \Delta^* & 0 & 0 & -U_k
  \end{array}\right),
\end{equation}
where the single particle energy is $U_k=\epsilon_k-\mu+T_kn$, the gap
is $\Delta=Up+g\,\phi_m$, and $\mu$ is the chemical potential. Here,
without cost, we have upgraded the mean-field contribution to the
single particle energy (which would otherwise be given by $Un$) to the
full ladder sum, $T_kn$, where $T_k$ is the two-body $T$-matrix. This
expression for the mean-field contribution to the particle energy is
an approximation which is accurate for dilute Fermi gases where the
quantum Fermi pressure limits achievable densities.  The dynamical
equations are closed by the evolution equation for the boson mode;
\begin{equation}
  i\hbar\frac{d\phi_m}{dt}=\nu\,\phi_m+ g\,p\,.
\label{phimeqn}
\end{equation}
The self-energy $\Sigma$ is diagonalized locally at each
$\boldvec k$ by the Bogoliubov transformation generating
quasiparticles with energy spectrum $E_k=\sqrt{U_k^2+|\Delta|^2}$. In
equilibrium, the quasiparticle states are occupied according to the
Fermi-Dirac distribution
$n_k=\left[\exp\bigl((E_k-\mu)/k_bT\bigr)+1\right]^{-1}$.  The
corresponding maximum entropy solution for the molecule amplitude is
found by $i\hbar d\phi_m/dt=\mu_m\phi_m$ where $\mu_m=2\mu$, so that
Eq.~(\ref{phimeqn}) implies $\phi_m=g\,p/(\mu_m-\nu)$. The mean fields
can then be determined by integration of the equilibrium single
particle density matrix elements, given by;
\begin{eqnarray}
  n&=&\frac1{(2\pi)^2}\int_0^Kdk\, \bigl[(2n_k-1)\cos2\theta_k
  +1\bigr]\,,\nonumber\\
  p&=&\frac1{(2\pi)^2}\int_0^Kdk\, (2n_k-1)\sin2\theta_k\,,
\label{eqfp}
\end{eqnarray}
where $\tan2\theta_k=|\Delta|/U_k$ is the Bogoliubov transformation
angle. Since $\theta_k$ depends on $n$ and $p$, these equations
require self-consistent solutions~\cite{pzero}.

So far this theory has been presented for a homogeneous system, while
we are interested in a gas of $N$ atoms confined in an external
trapping potential $V(\boldvec r)$. However, a full quantum mechanical
treatment of the trapping states is not required. For instance, in our
case, with a temperature of $T= 0.2\,T_F$, the harmonic oscillator
level spacing is smaller than both the Fermi and thermal energies.
Under these conditions, we may incorporate the effect of the trap
through a semiclassical local density approximation~\cite{SCA}.  This
involves replacing the chemical potential by a local one $\mu(\boldvec
r )=\mu-V(\boldvec r)$, and determining the thermodynamic solution at
each point in space as for the homogeneous system.

In general, the validity of the semiclassical approximation requires a
slow variation in the occupation of the discrete quantum levels as a
function of energy.  Remarkably, in both bosonic and fermionic gases,
this condition can often be satisfied even at very low temperatures;
because of strong correlations in a BEC due to repulsive interactions,
and because of exchange effects in a quantum Fermi gas. In both cases
the zero-temperature semiclassical approximation for dilute gases is
usually referred to as the Thomas-Fermi approximation.

\begin{figure}[h]
\begin{center}\
  \epsfysize=60mm \epsfbox{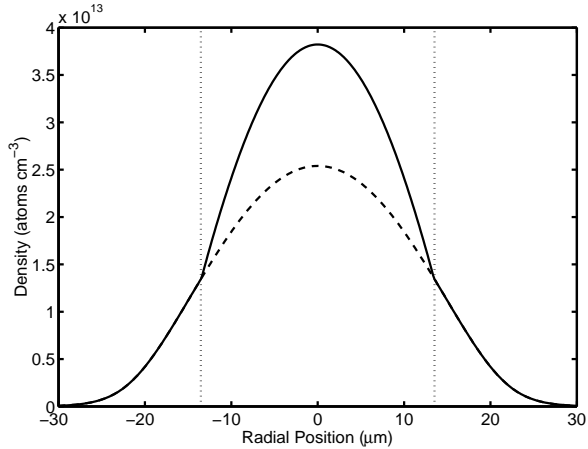}
\end{center}
\caption{
  Density profile at temperature $T=0.2\,T_F$ and detuning
  $\nu_0=20\,E_F$ showing accumulation of atoms at the trap center
  (solid line). We compare with the profile resulting from the same
  $\mu$ but artificially setting the pairing field $p$ to zero so that
  no superfluid is present (dashed line).}
\label{fig2}
\end{figure}

We evaluate the thermodynamic quantities at given $T$ and $N$ in three
steps: (i) For given $\mu$, we determine the local chemical potential
$\mu(\boldvec r)=\mu-V(\boldvec r)$ and use this value to find the
self-consistent solution for the density $n(\boldvec r)$ and pairing
field $p(\boldvec r)$ at each point in space, according to the
solution of Eqns.~(\ref{eqfp}); (ii) We modify the global chemical
potential $\mu$ until the density integral is the desired atom number,
i.e.  $N=\int d^3r\,n(\boldvec r)$; (iii) We use the resulting
solution for $\mu$ to calculate observable quantities, such as the
density, gap, compressibility, and so forth.

\begin{figure}[t]
\begin{center}\
  \epsfysize=60mm \epsfbox{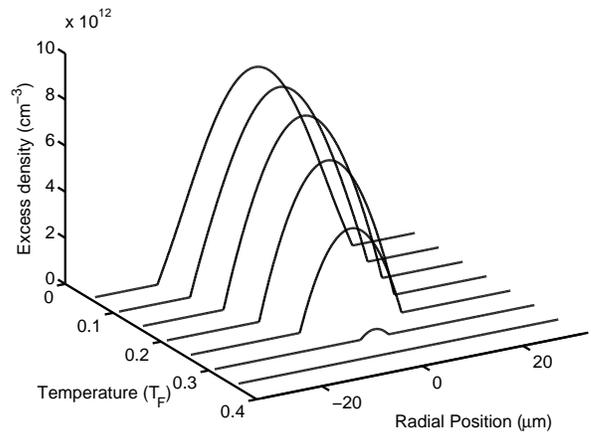}
\end{center}
\caption{
  Emergence of the coherent superfluid for $\nu_0=20\;E_F$. The
  superfluid occupies an increasing volume as the temperature is
  reduced. Shown is the excess density (difference between the dashed
  and solid lines in Fig.\protect{\ref{fig2}}) at each temperature.}
\label{fig3}
\end{figure}

The resulting solution for the density distribution is illustrated in
Fig.~\ref{fig2}. A striking signature of the resonance superfluidity
is evident in the predicted density profile which has a notable bulge
in the region of the center of the trap.  Experimentally, this
signature appears to be directly accessible. A typical approach would
be to fit the expected density profile for a quantum degenerate gas
with no superfluid phase to the wings of the distribution (outside the
dotted lines shown in Fig.\ref{fig2}). Then the excess density
observed at the trap center can be recorded. Fig.\ref{fig3}
illustrates the emergence of the superfluid as the temperature is
decreased. Qualitatively this situation is reminiscent of the central
condensate peak observed for a Bose-Einstein condensed gas in a
harmonic potential, although the connection appears to be somewhat
serendipitous.

\begin{figure}[h]
\begin{center}\
  \epsfysize=60mm \epsfbox{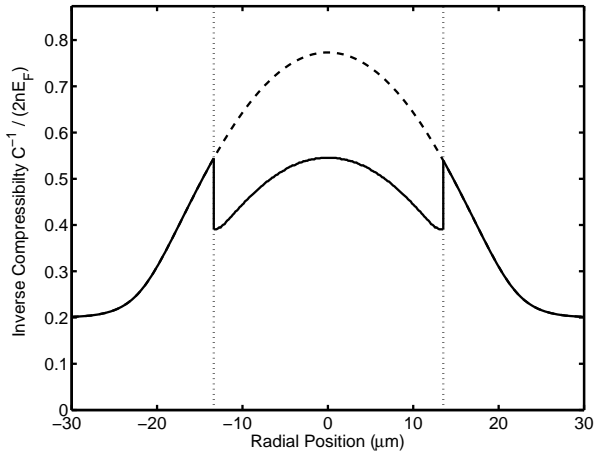}
\end{center}
\caption{Inverse isothermal compressibility $C^{-1}$ in units of the 
  Fermi energy (solid line). Here $\nu_0=20\,E_F$ and $T=0.2\,T_F$ (as
  can be seen from the limiting behavior at large radial position).
  A discontinuity appears at the radius at which the superfluid
  emerges (dotted line).  We compare this solution to that
  corresponding to zero pairing field and no superfluid phase
  transition (dashed line).  }
\label{fig4}
\end{figure}

We explain the observed behavior by considering the compressibility of
the normal and superfluid gas. Thermodynamically, the isothermal
compressibility $C$ is defined as $C^{-1}=n(\partial P/\partial n)_T$
where $P$ is the pressure, and is shown in Fig.~\ref{fig4}. The
compressibility is positive everywhere, indicating that, in spite of
the large attractive interactions, the Fermi pressure makes the
configuration mechanically stable. A significant feature is the
discontinuous behavior at the radius from the trap center at which the
superfluid changes from a zero to a non-zero value. This discontinuity
is a manifestation of a second-order phase transition occurring in
space. The discontinuity is a consequence of the local density
approximation, and cannot occur in a finite system. However, a rapid
change in the compressibility is expected. In principle this could be
probed by studies of shock waves generated by the abrupt jump in the
speed of sound as a density fluctuation passes through this region.

In conclusion, we demonstrated that there exists a direct signature of
superfluidity in trapped fermion gases. The onset of superfluidity
leads to a density bulge in the center of the trap which can be
detected by direct absorption imaging. The critical conditions for
superfluidity are satisfied initially in the trap center, and the
region of non-zero pairing field spreads out from the center as the
temperature is lowered further. The increase in the density profile in
the superfluid region is caused by a jump in the compressibility.
Direct measures of this behavior are possible by the study of the
propagation of sound waves.  We have applied our method here to
$^{40}$K, but a similar approach is easy to derive for other
interesting atoms, including in particular $^6$Li which is the other
fermionic alkali currently being investigated experimentally.

Support is acknowledged for M.H. and S.K. from the U.S. Department of
Energy, Office of Basic Energy Sciences via the Chemical Sciences,
Geosciences and Biosciences Division, and for M.C. from the National
Science Foundation and SNS, Pisa (Italy). We are grateful for the
hospitality of the Aspen Center for Physics where this work was
partially carried out. M.C. acknowledges the marvelous hospitality of
JILA, and thanks Ester Piegari for useful discussions.


\begin{references}
\bibitem{pzero} The $p=0$ solution should be discarded when the
  superfluid state is present since it then corresponds to an unstable
  point on the free energy surface.
\end{references}
\end{document}